\documentclass[a4paper,reqno,final]{amsart}
\usepackage{amsmath,amssymb,amscd,eucal}
\usepackage[T1]{fontenc}
\usepackage[latin1]{inputenc}
\usepackage{array}
\usepackage{epsfig}
\usepackage{mathrsfs}
\usepackage{bbold}
\usepackage[marginratio=1:1]{geometry}
\usepackage[notcite,notref]{showkeys}
\usepackage{hyperref}
\hypersetup{%
  pdftitle   = {Supersymmetry and spin structures},
  pdfkeywords = {supergravity, quotients, sphere, spin structure,
    supersymmetry},
  pdfauthor  = {José Figueroa-O'Farrill, Sunil Gadhia},
  pdfcreator = {\LaTeX\ with package \flqq hyperref\frqq}
}
\newcommand{\half}{\tfrac12}

\newcommand{\RR}{\mathbb{R}}
\newcommand{\ZZ}{\mathbb{Z}}
\newcommand{\1}{\mathbb{1}}
\newcommand{\fso}{\mathfrak{so}}
\newcommand{\fS}{\mathfrak{S}}
\newcommand{\Cl}{C\!\ell}
\newcommand{\RP}{\mathbb{RP}}
\newcommand{\bv}{\boldsymbol{v}}
\newcommand{\bw}{\boldsymbol{w}}
\DeclareMathOperator{\AdS}{AdS}
\DeclareMathOperator{\SO}{SO}
\DeclareMathOperator{\OO}{O}
\DeclareMathOperator{\Spin}{Spin}
\DeclareMathOperator{\dvol}{dvol}
\newcommand{\MUNCH}[1]{\relax}

\begin{document}

\title{Supersymmetry and spin structures}
\author[Figueroa-O'Farrill]{José Figueroa-O'Farrill}
\author[Gadhia]{Sunil Gadhia}
\address{School of Mathematics, The University of Edinburgh, Scotland,
United Kingdom}
\email{j.m.figueroa@ed.ac.uk, s.gadhia@sms.ed.ac.uk}
\begin{abstract}
  We construct examples of isometric M-theory backgrounds which
  preserve a different amount of supersymmetry depending on the choice
  of spin structure.  These examples are of the form $\AdS_4 \times
  L$, where $L$ is a seven-dimensional lens space whose fundamental
  group is cyclic of order $4k$.
\end{abstract}
\thanks{EMPG-05-14}
\maketitle

\section{Introduction}

The purpose of this brief note is to highlight and illustrate the
importance of specifying the spin structure as part of the data
defining a supergravity background.  For every positive integer $k$ we
will construct pairs of M-theory backgrounds with the same geometry
and flux but with two different spin structures and such that the
amount of supersymmetry which is preserved depends on the choice of
spin structure.  We will construct these backgrounds by quotienting
the Freund--Rubin background $\AdS_4 \times S^7$ \cite{FreundRubin} of
eleven-dimensional supergravity \cite{Nahm,CJS} by a cyclic group of
order $4k$ acting freely on the sphere.

The precise geometry of the Freund--Rubin background is
\begin{equation*}
  \AdS_4(-8s) \times S^7(7s) \qquad\text{with flux}\qquad F =
  \sqrt{6s} \dvol(\AdS_4)
\end{equation*}
where the parameter $s>0$ is the negative eleven-dimensional scalar
curvature $R = -s$ and the numbers in parenthesis are the scalar
curvatures of each of the geometries, this being the only modulus in a
manifold of constant sectional curvature.

The supersymmetry of this background boils down to the existence of
geometric Killing spinors in each factor:
\begin{equation*}
  \begin{split}
    \nabla_X \phi &= \tfrac16 f X \cdot \phi \qquad\text{on
      $\AdS_4$}\\
    \nabla_X \psi &= - \tfrac1{12} f X \cdot \psi \qquad\text{on
      $S^7$}~,
  \end{split}
\end{equation*}
where $f = \sqrt{6s}$.  The space of Killing spinors is
$32$-dimensional and as a representation of the isometry Lie algebra
$\fso(2,3) \oplus \fso(8)$ of the background it is isomorphic to
\begin{equation*}
  \fS^{2,3} \otimes \fS_-^{8,0}~,
\end{equation*}
where $\fS^{p,q}$ denotes the half-spin representation of $\fso(p,q)$
and the subscript denotes chirality, if relevant.  The chirality of
the $8$-dimensional half-spin representation has to do with the sign
of the ``Killing constant'' $f$ in the above Killing spinor equations.
Changing the sign of $f$ is tantamount to changing the sign of $F$
which in turn is tantamount to reversing the orientation of the
sphere.  We will use this device below to simplify the discussion in
certain points.  The lesson is thus not so much that the
$8$-dimensional spinor representation has negative chirality, but that
it \emph{is} chiral.

A cyclic group $\Gamma \subset \SO(8)$ acts isometrically on the round
sphere $S^7 \subset \RR^8$ by restricting to the sphere the linear
action on $\RR^8$.  We will assume that $\Gamma$ acts freely on the
sphere, so that no element (except the identity) fixes a point in the
sphere.  In this case, the quotient $S^7/\Gamma$ is smooth and locally
isometric to $S^7$: it is called a lens space.  In particular it has
constant positive sectional curvature, hence it is a spherical space
form.  The determination of all spherical space forms has a long
history culminating in Wolf's solution \cite{Wolf}.

As explained, for example, in \cite{FigSimAdS,FOMRS}, the quotient
$S^7/\Gamma$ will admit a spin structure if and only if the action of
$\Gamma$ on the orthonormal frame bundle of $S^7$ lifts to the spin
bundle.  The total space of the orthonormal frame bundle of the round
sphere $S^7$ is the Lie group $\SO(8)$, which fibres over $S^7$ with typical
fibre $\SO(7)$.  Indeed, $S^7$ can be thought of as the homogeneous
space $\SO(8)/\SO(7)$.  The action of $\Gamma \subset \SO(8)$ on
the orthonormal frame bundle is simply left multiplication in the
group $\SO(8)$ itself.

The total space of the spin bundle of $S^7$ is the
spin group $\Spin(8)$ which fibres over $S^7$ with fibre $\Spin(7)$.
We will let $\theta: \Spin(8) \to \SO(8)$ denote the two-to-one
covering map.  The action of $\Gamma$ on $\SO(8)$ will lift to the
spin bundle if and only if there exists a subgroup $\widehat\Gamma
\subset \Spin(8)$ which is mapped isomorphically to $\Gamma$ under
$\theta$.  The action of such $\widehat\Gamma$ on the spin bundle is
via left multiplication on the spin group itself.  Thus we see that
the spin structures in the quotient $S^7/\Gamma$ are in one-to-one
correspondence with the isomorphic lifts $\widehat\Gamma \subset
\Spin(8)$ of $\Gamma$.

A cyclic group $\Gamma$ is specified by exhibiting a generator $A$.
If $\Gamma$ has order $n$, then $A^n = 1$.  We will
investigate the existence of the subgroup $\widehat\Gamma \subset
\Spin(8)$ by lifting the generator $A$ to $\Hat A \in \Spin(8)$, there
being two such lifts distinguished by a sign, and then checking
whether there exists a choice of sign for which the relation $\Hat A^n
= 1$ is satisfied in $\Spin(8)$, thus recovering 
an isomorphic group.  We will work with $\Spin(8)$ inside the
Clifford algebra $\Cl(8)$, where in our conventions the Clifford
product obeys $\bv^2 = -|\bv|^2\1$ for $\bv\in\RR^8$, or in terms of
gamma matrices,
\begin{equation*}
  \gamma_i \gamma_j + \gamma_j \gamma_i = - 2 \delta_{ij} \1~.
\end{equation*}
The map $\theta:\Spin(8) \to \SO(8)$ is given explicitly in terms
of the Clifford algebra as follows.  Recall that $\Spin(8)$ embeds
in the Clifford algebra as the product of even number of elements of
$S^7 \subset \RR^8$:
\begin{equation*}
  \Spin(8) = \left\{ \bv_1 \bv_2 \cdots \bv_{2k} \mid  \bv_i \in
    \RR^8~,~|\bv_i|=1 \right\}~.
\end{equation*}
If $\bv \in \RR^8$ and $s = \bv_1 \bv_2 \cdots \bv_{2k} \in\Spin(8)$,
then
\begin{equation*}
  \theta(s) \cdot \bv = s \bv \Hat s~,
\end{equation*}
where $\Hat s = \bv_{2k} \cdots \bv_1$.  Since for $\bw \in S^7$,
$\bv \mapsto \bw \bv \bw = \bv - 2 \left<\bv,\bw\right> \bw$ is the
reflection in the hyperplane perpendicular to $\bw$, we see that this
formula exhibits the action of $\SO(8)$ on $\RR^8$ as a
composition of reflections.  In particular it is clear from this
observation that if $\bv\in\RR^8$ then so is $s\bv \Hat s$, as
claimed.

If $\Gamma \subset \SO(8)$ lifts isomorphically to $\widehat\Gamma
\subset \Spin(8)$, then we can investigate whether $S^7/\Gamma$ admits
any Killing spinors.  Bär's cone construction \cite{Baer} relates
Killing spinors on $S^7$ to parallel spinors on $\RR^8$, which are
themselves in one-to-one correspondence with the relevant half-spin
representation of $\Spin(8)$.  Moreover this correspondence is
equivariant with respect to the isometry group.  Therefore Killing
spinors on $S^7/\Gamma$ are in one-to-one correspondence with
$\widehat\Gamma$-invariant parallel spinors in $\RR^8$, or
equivalently with $\widehat\Gamma$-invariant spinors in the relevant
half-spin representation of $\Spin(8)$.  What this means in practise
is that we must check, for each isomorphic lift
$\widehat\Gamma$---that is, for each inequivalent spin structure in
the quotient---whether $\widehat\Gamma$ preserves any spinors in the
half-spin representations $\fS^8_\pm$.

\section{Seven-dimensional lens spaces}
\label{sec:7d}

Every cyclic subgroup $\Gamma\subset\SO(8)$ is conjugate (perhaps by
$\OO(8)$) to $\Gamma(n,a,b,c)$, a cyclic subgroup of order $n$
generated by
\begin{equation*}
  A = 
  \begin{pmatrix}
    R\left(\tfrac{1}{n}\right) & & & \\
    & R\left(\tfrac{a}{n}\right) & & \\
    & & R\left(\tfrac{b}{n}\right) & \\
    & & & R\left(\tfrac{c}{n}\right)
  \end{pmatrix}~,
\end{equation*}
where $R(\theta)$ denotes the rotation matrix
\begin{equation*}
  R(\theta) =
  \begin{pmatrix}
    \phantom{-}\cos 2\pi\theta & \sin 2\pi\theta\\
    - \sin 2\pi\theta & \cos 2\pi\theta
  \end{pmatrix}~,
\end{equation*}
and where $(a,n)=(b,n)=(c,n)=1$.  Without loss of generality we can
order them so that $1 \leq a \leq b \leq c < n$.  In choosing $a,b,c$
positive we may have used conjugation by $\OO(8)$.  Such conjugations
may change the orientation of the sphere, which in turn change the
Killing constant in the Killing spinor equation on $S^7$ or,
equivalently, the chirality of the parallel spinors in $\RR^8$. What
this means in practise is that we must consider \emph{both} half-spin
representations $\fS^8_±$.

There are two possible lifts of $A$ to $\Spin(8) \subset \Cl(8)$,
distinguished by a sign $\varepsilon$:
\begin{equation*}
  \Hat A = \varepsilon \exp\left( \frac{\pi}{n} \gamma_{12} +
    \frac{a\pi}{n} \gamma_{34} + \frac{b\pi}{n} \gamma_{56} +
    \frac{c\pi}{n} \gamma_{78} \right)~,
\end{equation*}
obeying
\begin{equation*}
  \Hat A^n = \varepsilon^n (-1)^{1+a+b+c} \1~.
\end{equation*}
We distinguish two cases.  If $n$ is even, then $a,b,c$ are odd and
hence $1+a+b+c$ is even.  Therefore $\Hat A^n = \1$ for either choice
of $\varepsilon$.  Therefore there are two inequivalent spin
structures in the corresponding quotient.  If $n$ is odd, we choose
$\varepsilon = (-1)^{1+a+b+c}$, whence there is a unique spin
structure in the quotient.

The eigenvalues of $\Hat A$ in the Clifford module are given by
\begin{equation*}
  \varepsilon \exp\left( \frac{i\pi}{n} \left( \sigma_1 + a \sigma_2
      + b \sigma_3 + c \sigma_4 \right) \right)
\end{equation*}
with $\sigma_i$ signs whose product $\sigma_1\sigma_2\sigma_3\sigma_4$
determines the chirality.  Since $1\leq a \leq b \leq c < n$, the only
way that this can be equal to $1$ is if
\begin{equation*}
  \sigma_1 + a \sigma_2 + b \sigma_3 + c \sigma_4 = 0, \pm n, \pm 2n
\end{equation*}
depending on the value of $\varepsilon$: $0,\pm 2 n$ for $\varepsilon
= 1$ and $\pm n$ for $\varepsilon = -1$.  In practise we do not have
to worry about this dichotomy, because the existence of an invariant
spinor implies the existence of a spin structure in the quotient, as
explained, for example, in \cite[Section~5.2]{FigSimAdS}.

Let us consider some cases as a way of illustration.

\subsection{$n=2$}

Here we have only one possible choice $a=b=c=1$, and the resulting
geometry is $\AdS_4 \times \RP^7$.  We will have an
invariant spinor whenever the weights $\sigma_i$ add up to $0,\pm2,\pm
4$.  For the ``positive'' spin structure, we require the sum to be
either $0$ or $\pm 4$.  This happens for the following choices of
weights: $\pm\pm\pm\pm$, $\pm\pm\mp\mp$, $\pm\mp\pm\mp$ and
$\pm\mp\mp\pm$, for a total of $8$ all of positive chirality.  For the
``negative'' spinor structure we require the sum to equal $\pm 2$.
This happens for $\pm\pm\pm\mp$, $\pm\pm\mp\pm$, $\pm\mp\pm\pm$ and
$\pm\mp\mp\mp$ for a total of $8$ all of negative chirality.  We
conclude that this quotient preserves \emph{all} of the supersymmetry
of the vacuum.

It was proved by Franc \cite{Franc} that of all the lens spaces,
$\RP^{4k+3}$ is the only one (apart from the sphere itself) admitting
the maximal number of Killing spinors and it was later proved by Bär
\cite{BaerSpheres} that this is still the case among all spherical
space forms.

\subsection{$n=3$}

The possible choices for $(a,b,c)$ are $(1,1,1)$, $(1,1,2)$ and
$(1,2,2)$.  The other choice $(2,2,2)$ gives the same quotient as
$(1,1,2)$ since they generate conjugate subgroups.  Let us take each
case in turn.

\subsubsection{$(1,1,1)$}

We require $\sigma_1 + \sigma_2 + \sigma_3 + \sigma_4 = 0$, which
happens for $6$ weights $\pm\pm\mp\mp$, $\pm\mp\pm\mp$ and
$\pm\mp\mp\pm$, all of positive chirality.

\subsubsection{$(1,1,2)$}

Here we require $\sigma_1 + \sigma_2 + \sigma_3 + 2 \sigma_4 = \pm 3$,
which happens for $6$ weights $\mp\pm\pm\pm$, $\pm\mp\pm\pm$,  and
$\pm\pm\mp\pm$, all of negative chirality.

\subsubsection{$(1,2,2)$}

Here we can have $\sigma_1 + \sigma_2 + 2 \sigma_3 + 2 \sigma_4 =
0,\pm 6$, which happens for $6$ weights $\pm\pm\pm\pm$, $\pm\mp\pm\mp$
and $\pm\mp\mp\pm$, all of positive chirality.

\subsection{$n=4$}

Again there are three possible choices for $(a,b,c)$: $(1,1,1)$,
$(1,1,3)$ and $(1,3,3)$, with $(3,3,3)$ and $(1,1,3)$ generating
conjugate subgroups.

\subsubsection{$(1,1,1)$}

Here we can have $\sigma_1 + \sigma_2 + \sigma_3 + \sigma_4 = 0, \pm
4$, with those weights adding up to $0$ and those to $\pm 4$ in
different spin structures.  For the positive spin structure, they must
add up to zero and there are six such weights of all positive
chirality: $\pm\pm\mp\mp$, $\pm\mp\pm\mp$ and $\pm\mp\mp\pm$.  For the
negative spin structure, they must add up to $\pm 4$ and there are two
such weights $\pm\pm\pm\pm$ all of positive chirality.

\subsubsection{$(1,1,3)$}

Again we can have $\sigma_1 + \sigma_2 + \sigma_3 + \sigma_4 = 0, \pm
4$.  For the positive spin structure, the sum must give $0$ which
happens for two negative-chirality weights: $\pm\pm\pm\mp$.  For the
negative spin structure, the sum must give $\pm 4$ which happens for
$6$ negative-chirality weights: $\mp\pm\pm\pm$, $\pm\mp\pm\pm$ and
$\pm\pm\mp\pm$.

\subsubsection{$(1,3,3)$}

Here we can have $\sigma_1 + \sigma_2 + \sigma_3 + \sigma_4 = 0, \pm
4, \pm 8$.  For the positive spin structure the sum must either be $0$
or $\pm 8$, which happens for $6$ positive-chirality weights:
$\pm\pm\pm\pm$, $\pm\mp\pm\mp$ and $\pm\mp\mp\pm$.  For the negative
spin structure, the sum must be $\pm 4$ and this happens for two
positive-chirality weights: $\pm\pm\mp\mp$.

Either one of these quotients constitutes possibly the simplest
example of the phenomenon which we would like to illustrate: the same
geometry $\AdS_4 \times (S^7/\ZZ_4)$ preserves a different amount of
supersymmetry depending on the choice of spin structure, in this case
either $\frac14$ or $\frac34$.

\subsection{$n=4k \geq 8$}

For $n=4k$, $k>1$, it is easily seen that the quotients $S^7/\ZZ_{4k}$
with weights $(a,b,c)$ given by $(1,2k-1,2k-1)$, $(1,2k+1,2k+1)$,
$(2k-1,2k+1,4k-1)$ have four invariant spinors with respect to the
positive spin structure and two with respect to the negative spin
structure, the chiralities being the same in both cases.   Similarly,
the quotient with weight $(1,2k-1,2k+1)$ has two invariant spinors
with respect to the positive spin structure and four with respect to
the negative spin structure, again with chiralities agreeing. 
We conclude that the corresponding supergravity backgrounds $\AdS_4
\times (S^7/\ZZ_{4k})$ are either $\half$-BPS or $\frac14$-BPS,
depending on the spin structure.

\subsubsection{$(1,2k-1,2k-1)$}

In this case the spinors with weights $\pm\mp\pm\mp$ and
$\pm\mp\mp\pm$ are invariant relative to the positive spin structure,
whereas the spinors with weights $\pm\pm\pm\pm$ are invariant relative
to the negative spin structure.  All have positive chirality.

\subsubsection{$(1,2k+1,2k+1)$}

In this case the spinors with weights $\pm\mp\pm\mp$ and
$\pm\mp\mp\pm$ are invariant relative to the positive spin structure,
whereas the spinors with weights $\pm\pm\mp\mp$ are invariant relative
to the negative spin structure.  All have positive chirality.

\subsubsection{$(2k-1,2k+1,4k-1)$}

In this case the spinors with weights $\pm\pm\pm\pm$ and
$\pm\mp\mp\pm$ are invariant relative to the positive spin structure,
whereas the spinors with weights $\pm\pm\mp\mp$ are invariant relative
to the negative spin structure.  All have positive chirality.

\subsubsection{$(1,2k-1,2k+1)$}

In this case the spinors with weights $\pm\pm\pm\mp$ are invariant
relative to the positive spin structure, whereas the spinors with
weights $\pm\mp\pm\pm$ and $\mp\pm\pm\pm $ are invariant relative 
to the negative spin structure.  All have negative chirality.

Moreover some experimentation suggests that these are (up to 
conjugation) the only cases where this phenomenon occurs.

\section{Conclusions and summary}
\label{sec:conc}

We have highlighted the importance of specifying the spin structure of
the spacetime as part of the data defining a supergravity background
by constructing examples of isometric M-theory backgrounds admitting
more than one spin structure and preserving a different amount of
supersymmetry depending on this choice.  Our examples are products of
$\AdS_4$ with lens spaces $S^7/\ZZ_{4k}$.  For $k=1$ the two
backgrounds are, respectively, $\tfrac14$- and $\frac34$-BPS, whereas
for $k>1$ they are $\frac14$- and $\half$-BPS, respectively.  We
expect this phenomenon to persist for other backgrounds which are
products of $\AdS_4$ with a spherical space form.  A systematic
analysis of such backgrounds is under way and we will be reporting on
these results in a more extensive forthcoming paper \cite{FigGadS7}.

\section*{Acknowledgments}

We are grateful to Elmer Rees for useful conversations.  JMF would
like to thank the IHÉS for hospitality and support during the time it
took to complete this work, and in particular Jean-Pierre Bourguignon
for the invitation to visit.  The research of SG is funded by a PPARC
Postgraduate Studentship.

\bibliographystyle{utphys}
\bibliography{AdS,Geometry,Sugra}

\end{document}